\documentclass[9pt,twoside]{rmaa-rho}
\RMxAAtemplatetype{\RMxAA}

\vol{100}
\pages{1000-1025}
\thisyear{2026}
\doi{\href{https://doi.org/10.22201/ia.01851101p.20XX.XX.XX.XX}{https://doi.org/10.22201/ia.01851101p.20XX.XX.XX.XX}}

\title{The Bimodal Mass Ratio Distribution of the Hyades}

\author[1]{Henri M.J. Boffin \orcidlink{0000-0002-9486-4840}}

\affil[1]{European Southern Observatory, Karl-Schwarzschild-str. 2, 85748 Garching, Germany}

\leadauthor{Henri M.J. Boffin}
\smalltitle{Deconvolving the Hyades}

\corres{Henri M.J. Boffin}
\email{hboffin@eso.org}

\received{April 30, 2026}
\accepted{}

\license{Texto de la licencia aquí}

\setbool{rho-abstract}{true} 
\setbool{rho-resumen}{false} 

\begin{abstract}
    The distribution of stellar mass ratios, $q$, in binary systems provides critical insights into the dynamical history and star-formation processes of open clusters. Here, I re-evaluate the recently published mass ratio distribution (MRD) of a sample of spectroscopic binaries in the Hyades cluster, which was based on stellar positions in colour-magnitude diagrams. I demonstrate that the  mass ratios derived in that work are statistically inconsistent with a random distribution of orbital inclinations. Furthermore, several systems yielded non-physical results. By applying a Richardson-Lucy deconvolution to the spectroscopic mass functions and assuming a random distribution of inclinations, I re-derive the MRD for this sample, showing that it is statistically different from a uniform distribution. I further find a significant dependence on the primary mass: systems with lower-mass primaries exhibit a peak near $q\approx1$, whereas more massive primaries show a distribution heavily skewed toward low mass ratios ($q\approx0.15$). These findings highlight the potential pitfalls of photometric mass ratio derivations and underscore the need for further verification with future data releases such as \textit{Gaia} DR4. 
\end{abstract}


\begin{resumen}
    La distribución del cociente de masas, $q$, en sistemas binarios proporciona información crítica sobre la historia dinámica y los procesos de formación estelar en cúmulos abiertos. En este trabajo, reevalúo la distribución del cociente de masas (MRD) recientemente publicada de una muestra de binarias espectroscópicas en el cúmulo de las Híades, que se basó en las posiciones estelares en diagramas color-magnitud. Demuestro que los cocientes de masa derivados en ese estudio son estadísticamente inconsistentes con una distribución aleatoria de inclinaciones orbitales, $i$. Además, varios sistemas arrojaron resultados no físicos. Al aplicar una deconvolución de Richardson-Lucy a las funciones de masa espectroscópicas, $f(M)$, y asumiendo una distribución aleatoria de inclinaciones, rederivo la MRD para esta muestra, mostrando que es estadísticamente diferente de una distribución uniforme. Además, encuentro una dependencia significativa con la masa primaria​: los sistemas con primarias de menor masa presentan un pico cerca de $q \approx 1$, mientras que las primarias más masivas muestran una distribución fuertemente sesgada hacia cocientes de masa bajos ($q \approx 0.15$). Estos hallazgos destacan los posibles errores en las derivaciones fotométricas del cociente de masas y subrayan la necesidad de una mayor verificación con futuras publicaciones de datos, como \textit{Gaia} DR4.
    
\end{resumen}

\begin{document}

\maketitle
\pagestyle{fancy}\thispagestyle{firststyle}


\section{BINARIES IN OPEN CLUSTERS}

At least half of all solar-like stars are found in binary or multiple systems, with this fraction rising to nearly 100\% among the most massive stars. Over their lifetimes, a significant number of these systems will undergo dynamical or evolutionary interactions -- processes that can profoundly reshape the structure and fate of both stellar components. Such interactions give rise to a diverse array of exotic objects that defy explanation by standard single-star evolution models. These include gravitational wave progenitors, blue stragglers, symbiotic and barium stars, novae, and certain types of supernovae. Beyond their intrinsic astrophysical interest, binary stars serve as indispensable laboratories: they constrain models of stellar evolution, star formation, and even the nature of gravity itself. Moreover, they provide a robust, model-independent means of determining fundamental stellar parameters -- masses, radii, and luminosities -- anchoring our understanding of stars across the cosmos \citep{2025CoSka..55c..21B}. 

Studying binary stars within open clusters offers a unique and powerful window into both stellar evolution and the dynamical history of star-forming environments. Open clusters, with their well-defined ages, metallicities, and distances, provide a controlled setting in which the properties of binary systems, such as mass ratios, orbital periods, and interaction frequencies, can be measured with amazing precision. By comparing the observed characteristics of binaries in clusters of different ages, astronomers can directly probe how dynamical processes (e.g., tidal interactions, mass transfer, and stellar mergers) shape the evolution of stars over time. Furthermore, the high stellar densities in young clusters promote the formation and early evolution of binary systems, making these environments ideal laboratories for testing theories of star formation and the origin of exotic objects like blue stragglers and compact object binaries. Ultimately, such studies not only refine our models of stellar and binary evolution but also shed light on the broader astrophysical processes that govern the life cycles of stars in the Universe \citep{2007A&A...473..829M,2017A&A...597A..68V,2025ARA&A..63..467M}.

In this respect, any extensive studies of the binary systems in open clusters is most welcome \citep[e.g.,][]{2026A&A...706A..62M}. One of these is the impressive work by 
\citet[][TSL26 in the following]{2026ApJS..283...81T}, which presents the results of more than 45 years of monitoring of the Hyades cluster. Building on previous work by Roger Griffin \citep{2012JApA...33...29G}, TSL26 presents the binary population of the closest open cluster to us. Among others, it presents the distribution of mass ratios of their sample of binaries, members of the cluster. In total, there are 54 single-lined binaries and 39 double-lined binaries.

       \begin{figure}[htbp]
            \centering
            \includegraphics[width=0.99\columnwidth]{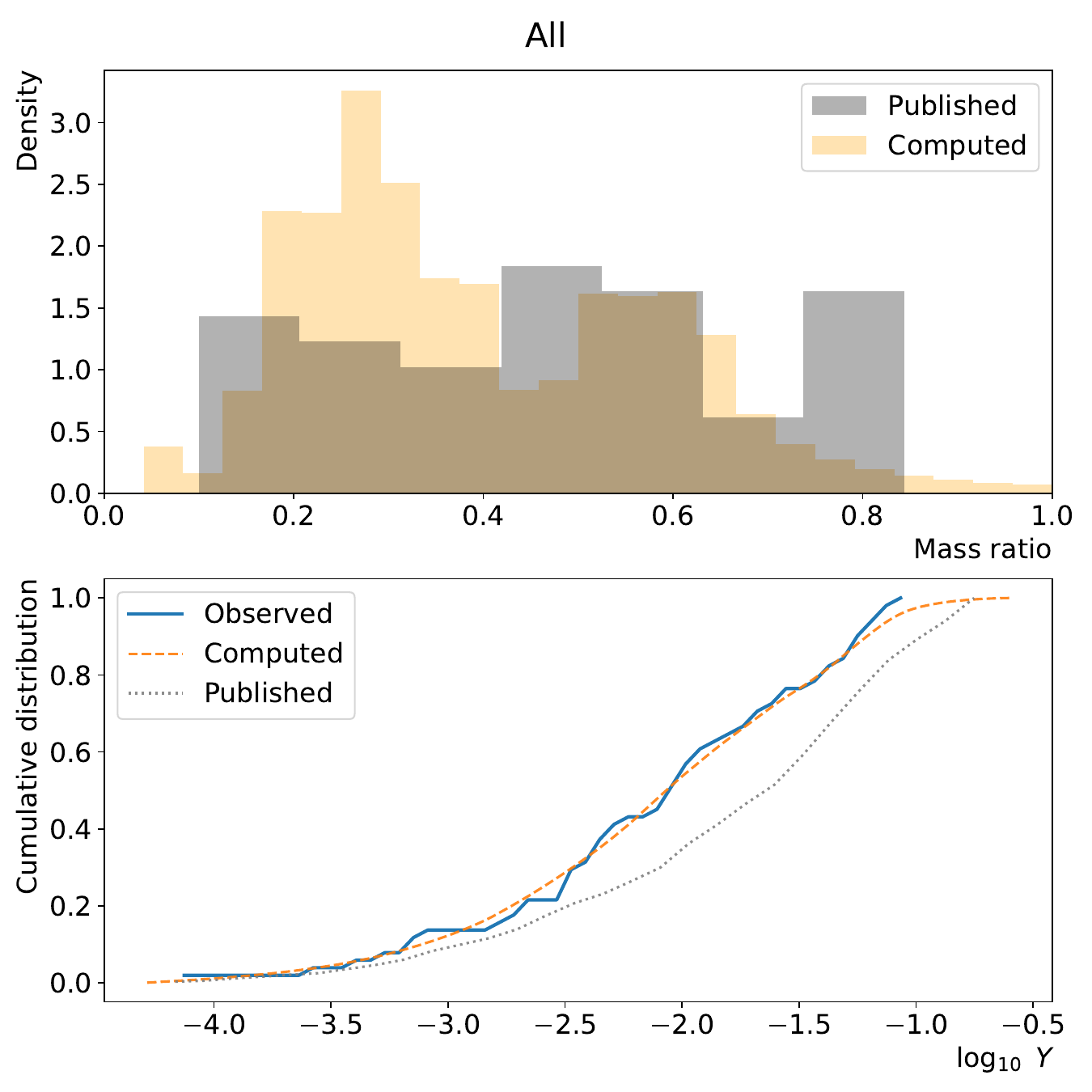}
            \caption{Top panel: Mass ratio distribution (MRD) for the SB1s of the TSL26 sample. The gray histogram shows the published MRD, while the orange histogram is the one I computed through deconvolution, assuming a random inclination on sky. Bottom panel: Associated cumulative distribution of $\log Y$ corresponding to the above MRDs, compared to the observed distribution.}
            \label{fig:figure1}
        \end{figure}

\section{THE MASS RATIO DISTRIBUTION}

As also mentioned by TSL26, a well-known feature of single-lined binaries (SB1) is that the mass ratio, $q$, cannot be directly derived from the spectroscopic orbit, as the orbital inclination on the sky, $i$, is apriori unknown (one would know it if the binary was also an eclipsing system or an astrometric binary). 
For SB1s, the orbit will provide the spectroscopic mass function, $f(m)$, given by
$$ f(m) = M_1 \frac{ q^3}{(1+q)^2} \sin^3 i,$$
where $M_1$ is the primary mass.

To remedy to this issue, one usually rely on statistical methods 
\cite[e.g.,][and refs. therein]{2010A&A...524A..14B,2017MNRAS.472.4497S,2020Obs...140....1B} that allow then to derive the mass ratio {\it distribution} (MRD). In order to avoid such a statistical analysis, and to be able to determine the mass ratio of each individual system, TSL26 made use instead of the fact that their binary systems are in an open cluster, that is, that their position in a colour-magnitude diagram (CMD) is well determined. They combined \textit{Gaia} and 2MASS photometry with their knowledge of the spectroscopic orbit to estimate the primary mass and the mass ratio of each binary system. This is in principle a very powerful method and widely used \citep{2023A&A...675A..89D,2025ApJ...989..104C,2026A&A...706A..62M}, but is, however, also potentially prone to  pitfalls. Indeed, differential extinction or errors on the parallax may displace the object in the CMD and thereby lead to an incorrect mass ratio derivation. The Hyades being so close and the extinction being quite small, the first effect is hopefully negligible. The second effect is more difficult to assess, unfortunately, as the binarity of the systems will likely affect the \textit{Gaia} astrometry and thus may lead to an incorrect parallax. This will hopefully be addressed in \textit{Gaia} DR4, but is still an issue in \textit{Gaia} DR3 as the majority of stars have been analysed as if they were single. Another possible hurdle is the fact that some of the binary systems may contain a non-luminous, degenerate object as the result of stellar evolution. In that case, the companion wouldn't be visible in the CMD and the derived mass ratio would be erroneous. As the binaries in open clusters could contain up to 25\% of systems harbouring a white dwarf\footnote{This fraction is likely much less for a young open cluster like the Hyades, where the turn-off mass is about 2.3 M$_\odot$.} \citep{2017A&A...597A..68V}, this may be a non-negligible issue. Finally, the reliance on theoretical isochrones to apply this method may also lead to issues \citep{2026arXiv260420722M}.

    \begin{figure}[htbp]
            \centering
            \includegraphics[width=0.99\columnwidth]{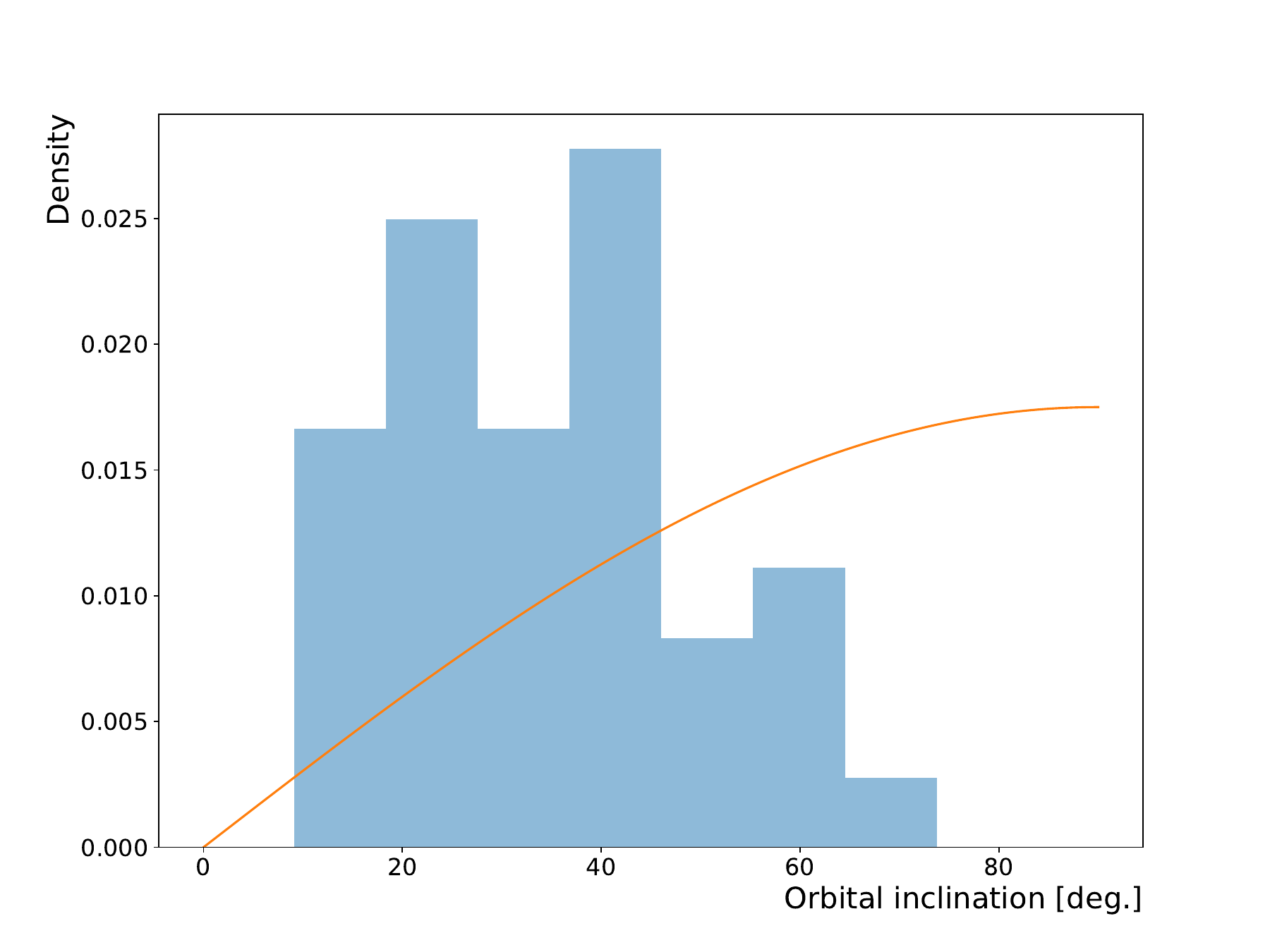}
            \caption{The distribution of orbital inclination in degrees associated to the MRD of TSL26 is shown as a blue histogram, while the expected random distribution is indicated with the orange solid line.}
            \label{fig:inc}
        \end{figure}

         \begin{figure*}[htbp]
            \centering
            \includegraphics[width=1.99\columnwidth]{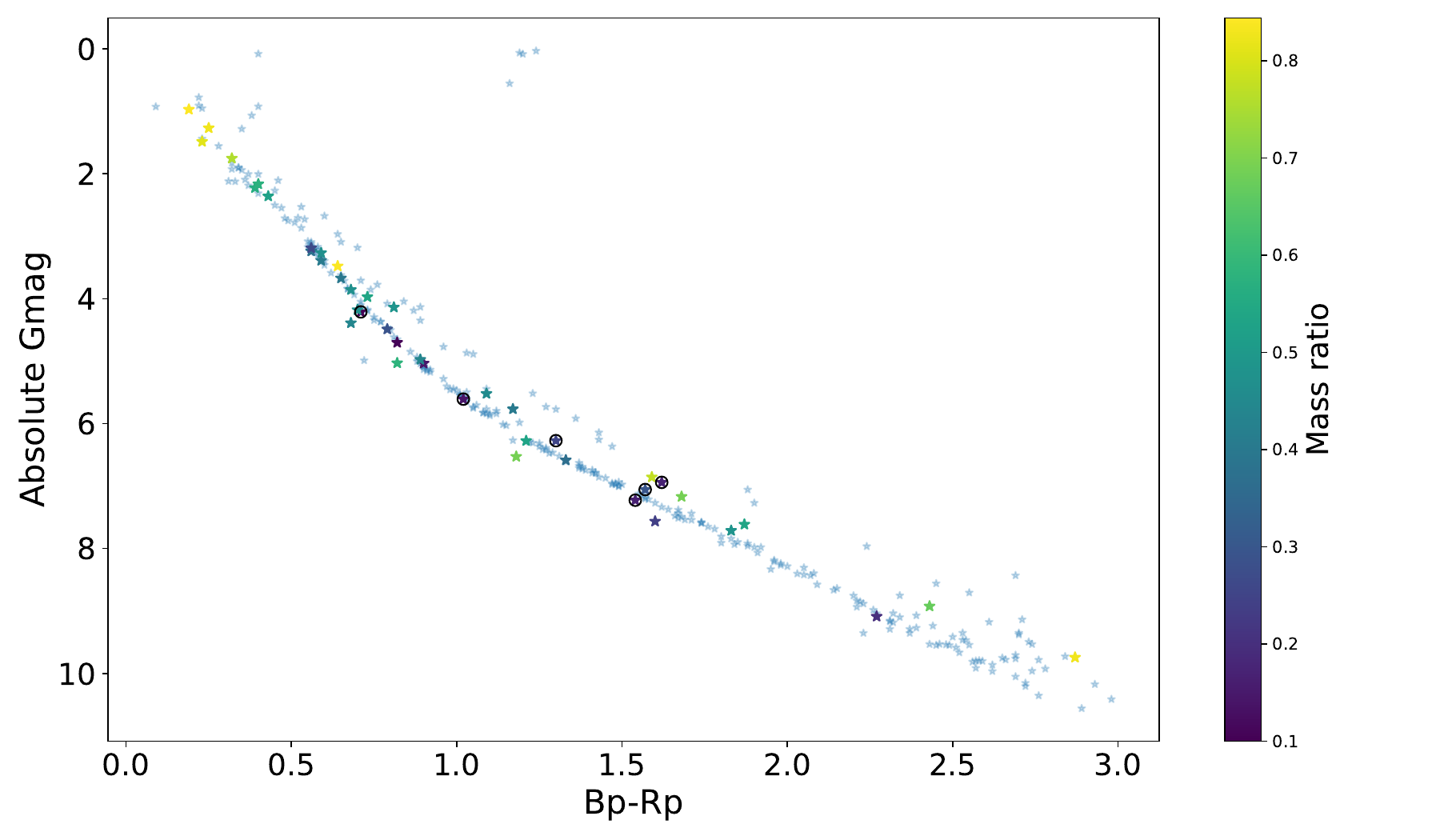}
            \caption{Gaia colour-magnitude diagramme of the Hyades members as determined by TSL26. The SB1 systems are coloured based on the mass ratio determined by the same authors. Six of the seven systems that lead to an nonphysical inclination angle are shown surrounded by a circle.}
            \label{fig:cmd}
        \end{figure*}
        
       \begin{figure}[htbp]
            \centering
            \includegraphics[width=0.99\columnwidth]{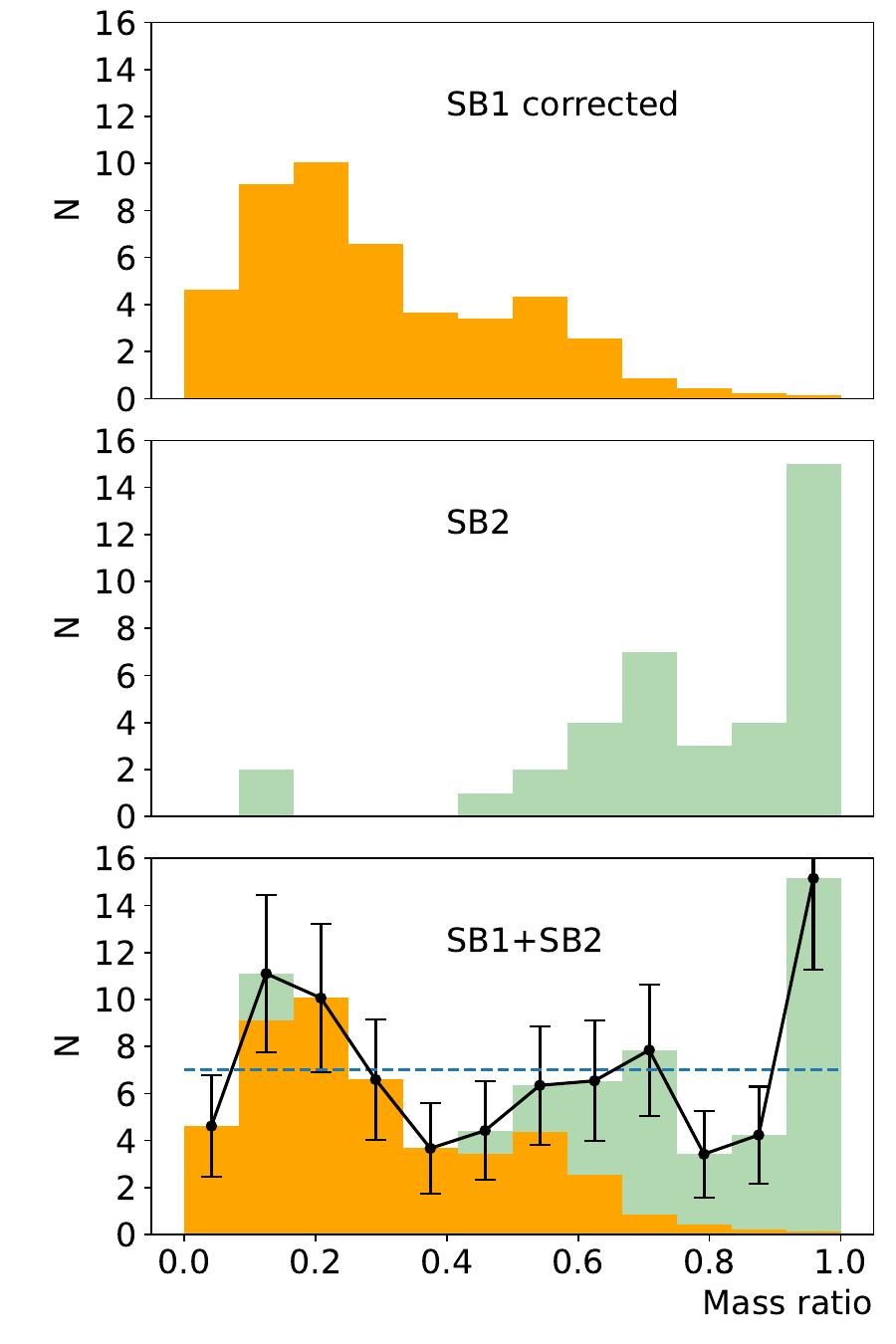}
            \caption{Mass ratio distributions (MRD) for the TSL26 sample. The top panel shows the incompleteness-corrected MRD for SB1, based on the results of Fig.~\ref{fig:figure1}. The middle panel presents the MRD for SB2s, while the bottom panel is the sum of the two above, and thus the final MRD.}
            \label{fig:figure2}
        \end{figure}

\subsection{A simple verification}
There is in principle an easy way to verify this. Indeed, {\it if we can assume that the orbital inclinations are randomly distributed} -- a quite natural assumption to make -- then we can use the derived mass ratio distribution to obtain the distribution of $Y = f(m)/M_1$ and compare it to the observed one. The two distributions should thus match.

I have done this for the mass ratio distribution obtained for the final sample of 46 SB1s for which TSL26 could derive the masses of the components. The distribution of $Y$ so derived, shown in the lower panel of Fig.~\ref{fig:figure1}, appears very different from the observed one. A Kolmogorov-Smirnov test, implemented in {\tt scipy}, confirms this: the $p$-value that the two distributions are extracted from the same one is 0.00373, well below the usual accepted 0.05 threshold. Thus, {\it the distribution of mass ratios derived by TSL26 is not compatible with a random distribution of orbital inclinations!} It this is confirmed, then this is already a very strong result However, 
unless there some good reasons to think that this is indeed the case, we may  wonder if the mass ratio distribution is not erroneous and affected by some biases.

To be more quantitative, it is possible to obtain from the values of $f(m)$, $M_1$, and $q$, the orbital inclination of a given system and compare its distribution with the expected random distribution, $g(i)~di = \sin i ~di$. This is done in Fig.~\ref{fig:inc} where it is clear that the distribution of computed inclinations is very far from what is expected, with a preponderance of low orbital inclinations. 
If the distribution of orbit inclinations appears non-uniform because of some observational bias, we would expect the higher inclinations to be more represented, contrarily to what we see. Indeed, the semi-amplitude of the radial velocity variations are proportional to $\sin i$, so higher inclinations are more easy to detect. Thus, it is more likely that there is some fundamental flaw in the method used by TSL26 to derive the mass ratios. This is corroborated by the fact
that for 7 SB1s, the derived mass ratio is incompatible with a value of $\sin i \leq 1$ (with only one case that could be solved given the error on the mass function), likely indicating some issues as this is not physical. These systems are listed in Table~\ref{tab:unphy}. It is not clear what these systems have special compared to the others, although it does seem that for six of the seven systems, the Gaia RUWE was rather high -- larger than 3 and up to 10 in some cases. This would indicate that the Gaia solution cannot be trusted. However, there are several other systems with such large RUWE in their sample -- not surprisingly as they are binaries with long periods. In fact, most of the SB1s from the paper of TSL26 have a RUWE above 1.4, indicating that their Gaia astrometry is not reliable. This may be an issue when trying to put them in a CMD.

Fig.~\ref{fig:inc} indicates a predominance of low orbital inclinations, which would indicate that the mass ratios derived are generally larger than what would be expected for a random distribution of inclinations. Fig.~\ref{fig:cmd} reveals that the mass ratios derived by TSL26 do not seem to correlate very well with the positions of the systems in the colour-magnitude diagrams. The exact reason why the method is failing in not clear, but is likely due to at least a combination of the flaws indicated in the previous section.

\begin{figure}[htbp]
            \centering
            \includegraphics[width=0.99\columnwidth]{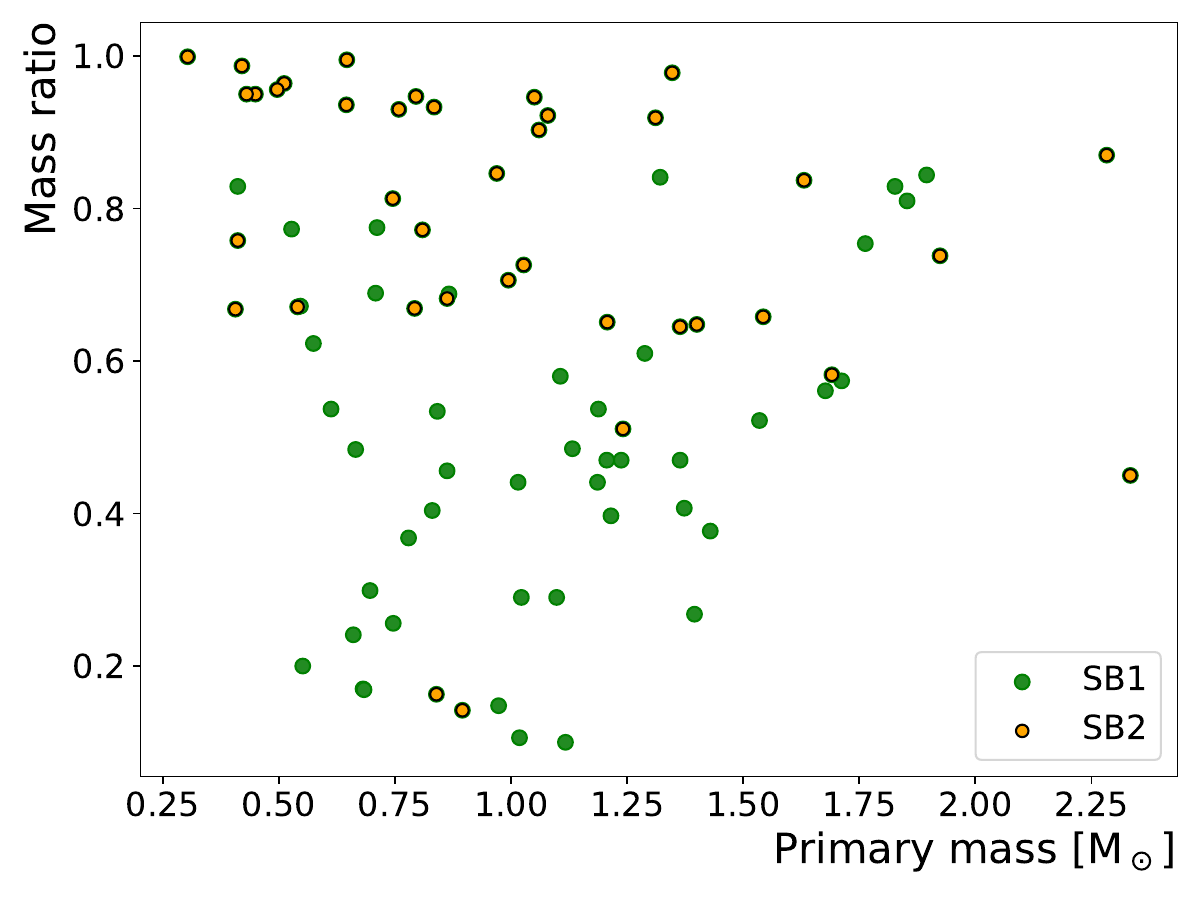}
            \caption{Mass ratios as determined by TSL26 as a function of their derived primary mass. The SB1 and SB2 are separated.}
            \label{fig:massq}
        \end{figure}

       \begin{figure}[htbp]
            \centering
            \includegraphics[width=0.99\columnwidth]{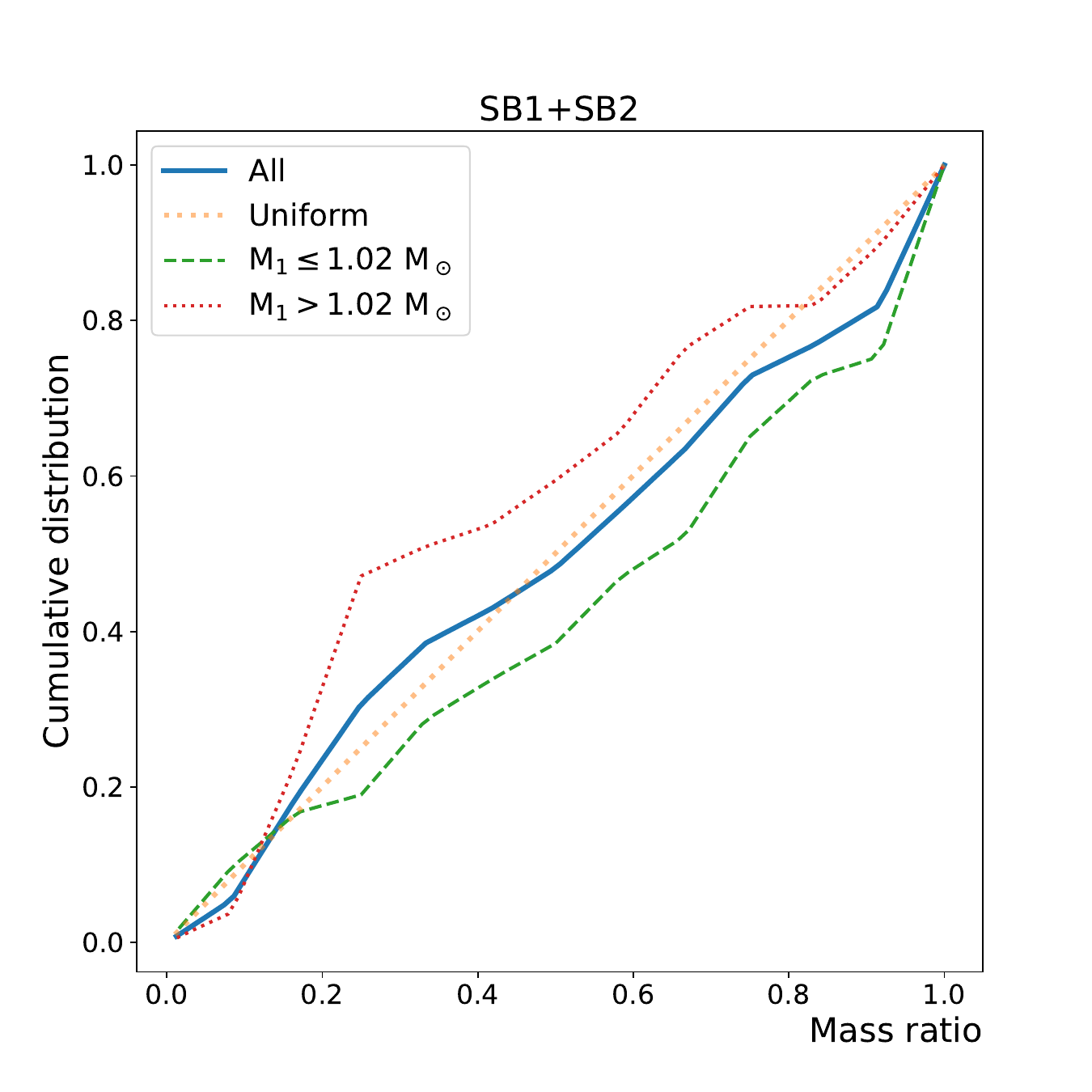}
            \caption{The normalised cumulative distribution of the mass ratios for the whole sample of spectroscopic binaries, based on Fig.~\ref{fig:figure2} is presented with the blue, solid line, while the sample split according to the primary mass is shown with the dashed and dotted lines. The uniform distribution is illustrated with the orange dotted line.}
            \label{fig:figure4}
        \end{figure}

\begin{figure*}[htbp]
        \centering                
        \includegraphics[width=0.49\textwidth]{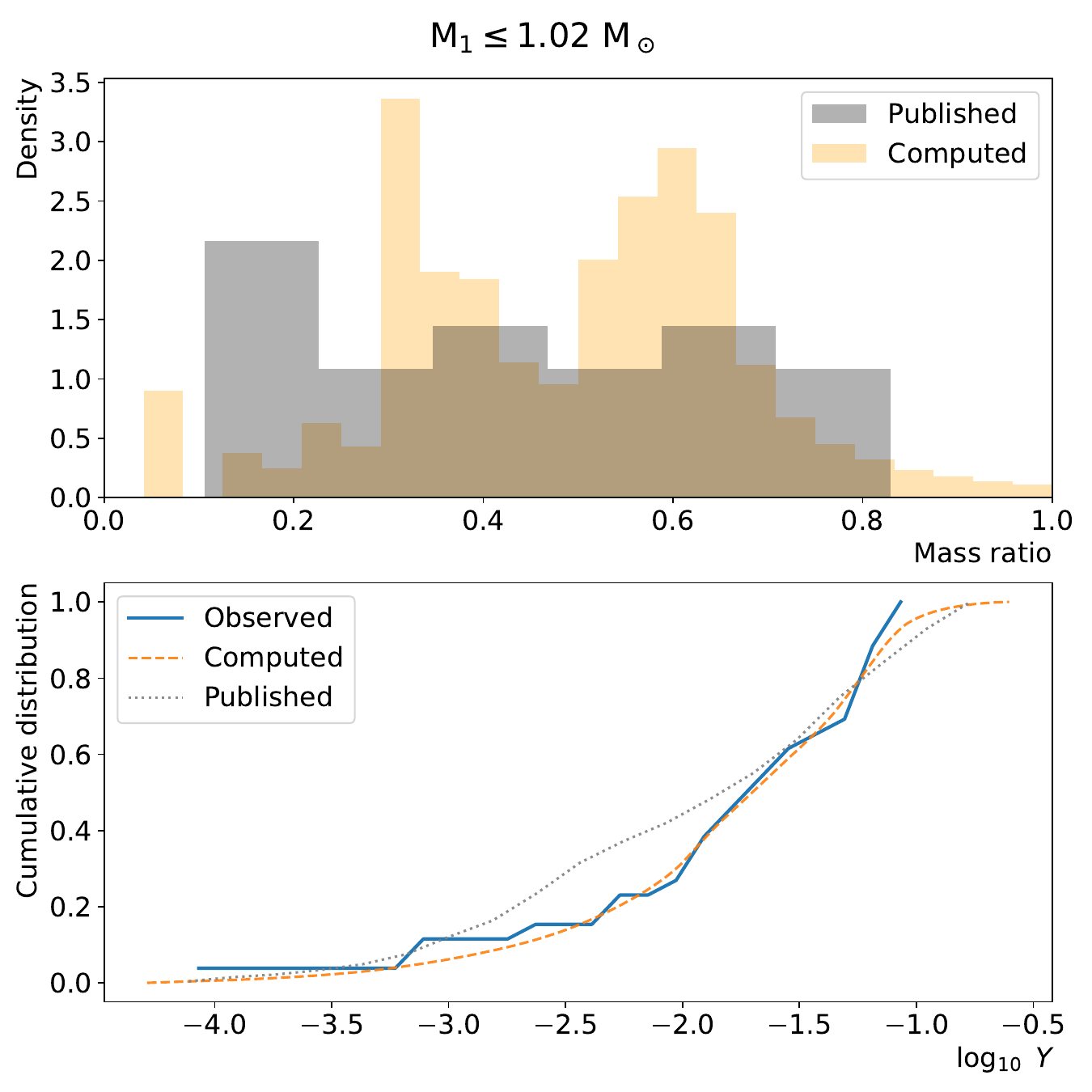}
        \includegraphics[width=0.49\textwidth]{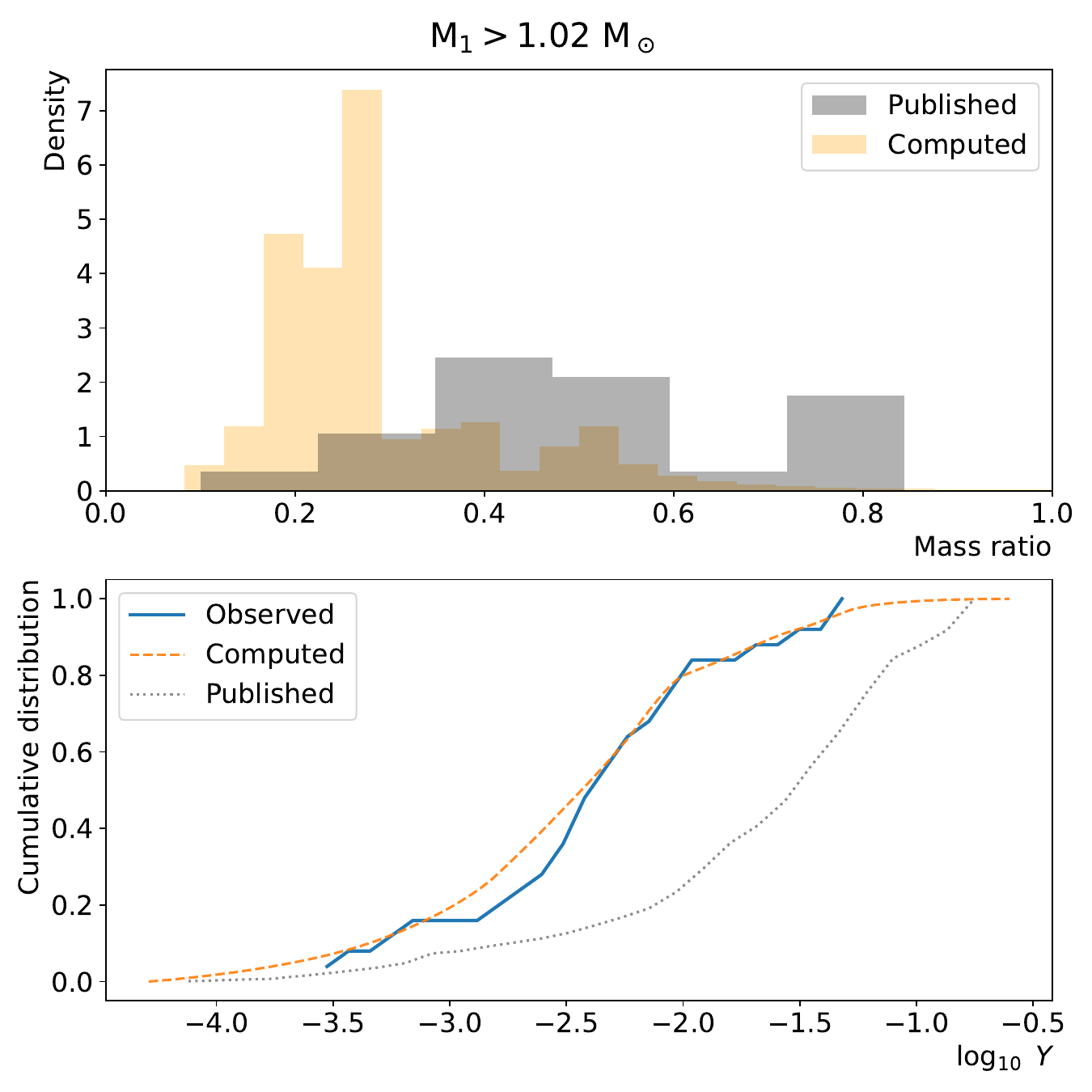}
        \caption{Same as Fig.~\ref{fig:figure1} in the case where the original sample is split in two, based on the primary mass. }
        \label{fig:mrall}
\end{figure*}

\begin{figure*}[htbp]
        \centering                
        \includegraphics[width=0.49\textwidth]{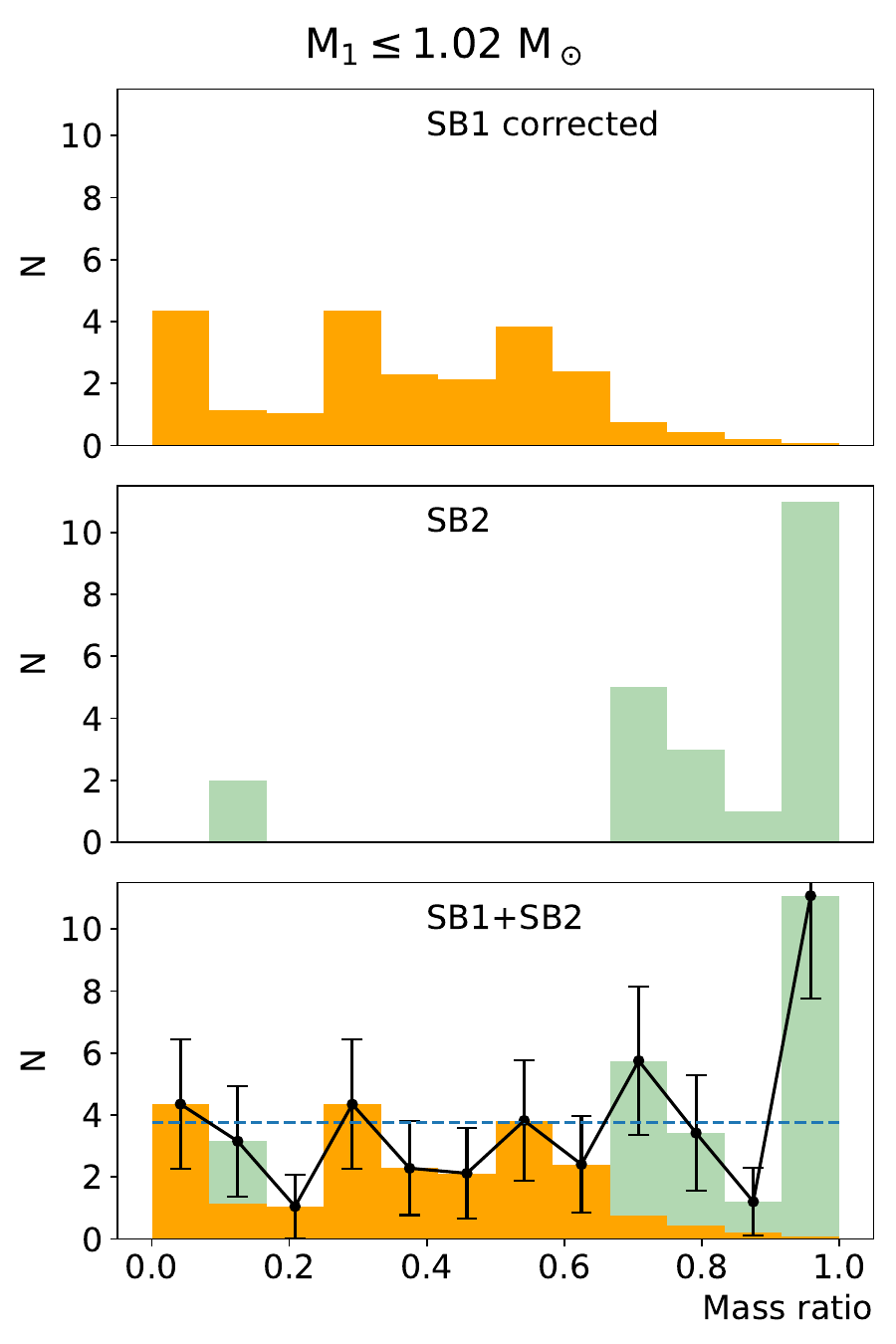}
        \includegraphics[width=0.49\textwidth]{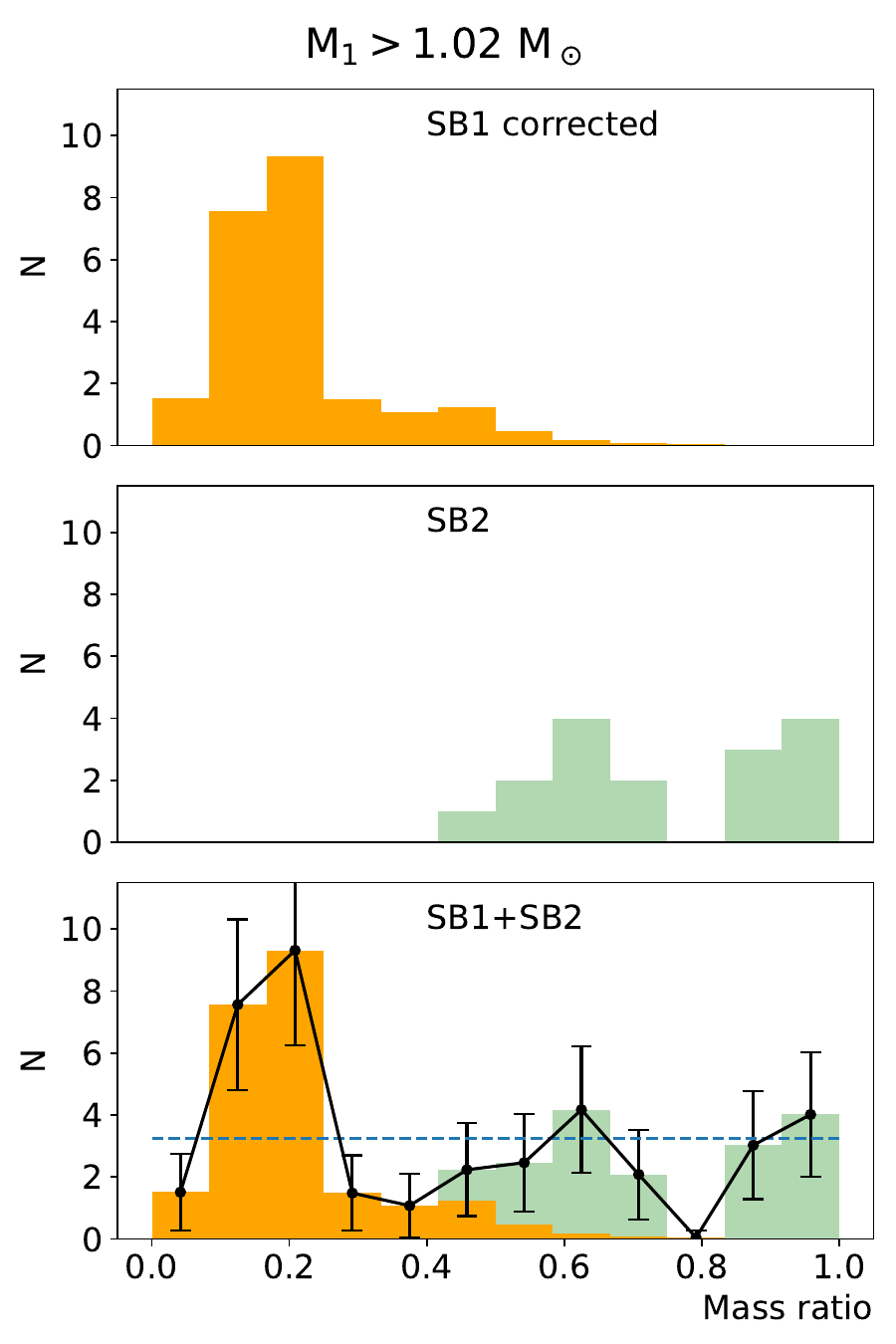}
        \caption{Same as Fig.~\ref{fig:figure2} in the case where the original sample is split in two, based on the primary mass. }
        \label{fig:mrcorsample}
\end{figure*}

\begin{table*}
\centering
\caption{The seven SB1 from TSL26 that lead to an nonphysical $\sin i > 1$. The ID, name, orbital period in days, eccentricity, mass ratio ($q$), mass function ($f(m)$) in M$_\odot$, primary mass ($M_1$), and the Gaia RUWE are indicated.
\label{tab:unphy}}
\begin{tabular}{llrllcrr}
\toprule
ID	&NAME		&Period	&ecc		&$q$	&$f(m)$	&$M_1$	& RUWE \\
\midrule
251	&vB 39	&4616.107299058	&0.847615	&0.29	&0.040444 $\pm$ 0.000837	&1.022	& \\
313	&vB 59	&5702.096505136	&0.940779	&0.1	&0.010292 $\pm$ 0.001009	&1.117	& 3.4\\	
337	&vB 275	&1915.180931688	&0.481505	&0.256	&0.039036 $\pm$ 0.001459	&0.746	& 10.6\\
425	&PELS 77	&2020.793441147	&0.515		&0.17	&0.047659 $\pm$ 0.001825	&0.681	& 9.9\\
506	&vB 115	&1203.190490385	&0.46705	&0.142	&0.014673 $\pm$ 0.000341	&0.895 & 7.6	\\
525	&StKM 2-390	&4975.432523114	&0.527591	&0.169	&0.007015 $\pm$ 0.004342	&0.683	& 8.9\\
591	&StKM 1-559	&112.181891598	&0.305528	&0.299	&0.028530 $\pm$ 0.000387	&0.696	& 5.6\\
\bottomrule
\end{tabular}
\end{table*}

\subsection{Rederiving the MRD}
We can now do the opposite exercise and assume that the orbital inclinations are randomly distributed on the sky. Given that the study of TSL26 is based on more than 45 years of monitoring, we can indeed assume that all the binaries that could be detected will be so. As such, there is no reason to think that the orbital inclinations are not randomly distributed.
I used the Richardson-Lucy deconvolution method \citep{1993A&A...271..125B,2020Obs...140....1B} to obtain from the distribution of $Y$ the mass-ratio distribution. The result is shown in Fig.~\ref{fig:figure1}, together with the recomputed distribution of $\log Y$. By construction, the latter will follow closely the observed distribution and indeed the associated $p$-value of the KS-test is 0.99.

Contrarily to TSL26, I cannot associate a given mass ratio to a given binary system: only the distribution can be obtained. The mass ratio distribution I obtain shows some differences with the published one. First, I do not obtain many large values of the mass ratios ($q > 0.7$), which is indeed appropriate for SB1, where there must be a large difference of flux between both components (and thus a large difference in mass) to not detect the secondary. On the other hand, my distribution shows a larger fraction of systems with low mass ratios, $q \approx 0.3$. My distribution agrees with TSL26's one as far as the low number of systems with very small mass ratios, but as shown by these authors, this is a selection effect as it is hard to detect a binary system where the companion is not very massive. 

The TLS26 distribution should indeed be multiplied by an incompleteness correction function in order to obtain the {\it true} distribution (see their Fig. 8). This is done in the top panel of Fig.~\ref{fig:figure2}, where one can see that the lower mass ratios are now preferred. The middle panel of this figure shows the MRD for the double-lined spectroscopic binaries (SB2). In this case, the mass ratio is a direct  outcome of the spectroscopic orbit determination as it is simply the ratio of the two semi-amplitudes of radial velocities. As expected, SB2s show a preference for higher mass ratios, with a peak in the last bin of $q \geq 0.95$. Such a preference for equal mass systems is a known, although still debated, feature of binary samples (see below). Here, such peak at $q \approx 1$ seems to arise from lower-mass stars, as revealed by Fig.~\ref{fig:massq} that shows the mass ratios as determined by TSL26 as a function of the primary mass. 

The final MRD would consist in summing the ones for SB1s and SB2s, as shown in the bottom panel of  Fig.~\ref{fig:figure2}, which assumes that there is no need to correct the SB2 MRD for any incompleteness factor as these are normally easily detected. The final distribution shows several peaks, at $q \approx 0.15, 0.7$, and 0.95, and despite the low-number statistics, indicates significant departure from a uniform distribution. This is indeed confirmed when plotting the normalised cumulative distribution (Fig.~\ref{fig:figure4}), which is different from the one expected for a uniform distribution. This is also confirmed by a K-S test, which provides a $p$-value undistinguishable from 0, that is, a very statistically significant difference.

\section{DEPENDENCE ON PRIMARY MASS}

In the same way that the binary fraction is a function of the mass of the primary star, it is now well established that the mass-ratio distribution is also dependent on the primary's mass \citep{2019MmSAI..90..359B}. It is therefore interesting to see if such an effect can be detected here as well. I have therefore split the TSL26 sample of SB1s into two equal groups of 23 systems, one with the primary mass below 1.02 M$_\odot$, which would correspond to stars of spectral types G-K-M, and the other with $M_1 > 1.02 M_\odot$, that is G-F-A spectral types. The median primary mass in the two samples is 0.7 M$_\odot$ and 1.3 M$_\odot$, respectively. 

I then apply the same strategy and derive the mass ratio distribution using the Richardson-Lucy method on both samples. The results are shown in Fig.~\ref{fig:mrall}. Here again, we clearly see differences with the published distribution for SB1s, with the effect most evident for the sample of the systems with a primary mass above 1.02 M$_\odot$. Quite surprisingly, for the sample of lower mass stars, I find a deficit of low mass ratios compared to what is published, while for the more massive stars, the distribution is very much peaked around low mass ratios, $q \approx 0.2$. In both cases, the resulting cumulative distributions of $\log Y$ are very different from the one published, with the latter not in agreement with the observed ones. 

I then proceed in the same way as for the whole sample and apply the completeness factor, and sum up the distributions of SB1s and SB2s, to derive the final distributions (Fig.~\ref{fig:mrcorsample}). This again shows the striking differences between the two subsamples, with a peak close to $q \approx 1$ for the lower mass systems, and a peak around $q \approx 0.15$ for the more massive ones. 

The peak at $q \approx 1$ is referred in the literature as an overabundance of twins, and its significance has been debated \citep{2006A&A...457..629L}. They are generally observed mostly in binary systems containing a G-type primary with short orbital periods, below 50 days \citep{2009AJ....137.3442S,2022ApJ...933..119L}. A nice interpretation of the peak at $q \approx 0.15$ would be that these corresponds to systems with white dwarf companions, as mentioned earlier. However, as the median value of the primary mass for the more massive sample is 1.3 M$_\odot$, such a peak corresponds to a companion mass of about 0.2 M$_\odot$, which is too small to be due to white dwarfs.

Following a useful suggestion from an anonymous referee, I also looked at a possible dependence of the MRD on the orbital period -- separating the sample in two, according to whether they have an orbital period below the median value for SB1s, which is about 800 days. I could not find significant differences between these two samples and none were showing an excess of twins or low mass ratios. With much larger samples that will be available hopefully with the Gaia DR4, such detailed studies could however reveal interesting features.

\section{A NOTE OF CAUTION}
In this paper, I have shown that the mass ratio distribution computed by TSL26 based on the position of the systems in the colour-magnitude diagram are not compatible with the assumption of a random distribution of orbital inclinations on the sky. The derived mass ratios do not seem to correlate with the value expected from their position in the CMD. Moreover, for seven objects, the mass ratio that was derived is not compatible with the associated primary mass and mass function, and are thus likely not correct. 

Assuming a random orbital inclination distribution, I have recomputed the mass ratio distribution and shown it to be statistically different from a uniform distribution. Moreover, splitting the systems according to the primary mass allows us to distinguish quite different features in the MRD. It will be interesting to see if \textit{Gaia} DR4 will confirm these findings. It will also be useful to extend this analysis to other open clusters and see if the MRD is showing any correlation with the cluster's age or total mass. 

\renewcommand{\refname}{REFERENCES}
\bibliographystyle{rmaa}
\bibliography{sample}

\end{document}